\documentclass[reprint,superscriptaddress,amsmath,amssymb,aps,prl]{revtex4-1}
\pdfoutput=1

\usepackage{graphicx}
\usepackage{dcolumn}
\usepackage{bm}
\usepackage{epstopdf}
\usepackage{hyperref}

\begin{document}

\title{Multithermal-Multibaric Molecular Simulations from a Variational Principle}

\author{Pablo M. Piaggi}
\affiliation{Theory  and  Simulation  of  Materials  (THEOS), {\'E}cole Polytechnique F{\'e}d{\'e}rale de Lausanne,  CH-1015  Lausanne,  Switzerland}
\affiliation{Facolt{\`a} di Informatica, Istituto di Scienze Computazionali, and National Center for Computational Design and Discovery of Novel Materials (MARVEL), Universit{\`a} della Svizzera italiana (USI), Via Giuseppe Buffi 13, CH-6900, Lugano, Switzerland}
\author{Michele Parrinello}%
 \email{parrinello@phys.chem.ethz.ch}
\affiliation{Department of Chemistry and Applied Biosciences, ETH Zurich, c/o USI Campus, Via Giuseppe Buffi 13, CH-6900, Lugano, Switzerland}
\affiliation{Facolt{\`a} di Informatica, Istituto di Scienze Computazionali, and National Center for Computational Design and Discovery of Novel Materials (MARVEL), Universit{\`a} della Svizzera italiana (USI), Via Giuseppe Buffi 13, CH-6900, Lugano, Switzerland}

\date{\today}

\begin{abstract}
We present a method for performing multithermal-multibaric molecular dynamics simulations that sample entire regions of the temperature-pressure (TP) phase diagram.
The method uses a variational principle [Valsson and Parrinello, Phys.\ Rev.\ Lett.\ \textbf{113}, 090601 (2014)] in order to construct a bias that leads to a uniform sampling in energy and volume.
The intervals of temperature and pressure are taken as inputs and the relevant energy and volume regions are determined on the fly.
In this way the method guarantees adequate statistics for the chosen TP region.
We show that our multithermal-multibaric simulations can be used to calculate all static physical quantities for all temperatures and pressures in the targeted region of the TP plane.
We illustrate our approach by studying the density anomaly of TIP4P/ice water.
\begin{description}
\item[PACS numbers]
05.10.-a, 31.15.xv, 05.20.Gg, 31.15.xt
\end{description}
\end{abstract}

\maketitle

Present day condensed matter studies are heavily dependent on atomistic modeling.
Hardly any paper, theoretical or experimental, appears that is not accompanied by some form of numerical modeling.
In many areas this implies performing a molecular dynamics (MD) or Monte Carlo simulations based on an atomistic description of matter.
This has spurred an intensive research effort whose objective has been to make simulations more accurate and efficient.  
One problem is the sheer computational cost of the simulation, this rises with system size and even more steeply with the accuracy with which the interactions are computed.
A case in point is \textit{ab initio} molecular dynamics. 
This calls for an efficient use of the simulation that allows obtaining  a maximum of information with minimum effort.

One avenue that has been followed to make better use of the sampling time has been to alter the probability with which states with different energies are sampled.
In standard simulations one samples the Boltzmann distribution in which the probability of observing rare energy fluctuations away from its average value is exponentially suppressed.
For this reason it has been suggested to replace Boltzmann sampling with a different one, in which a different energy distribution is sampled and  later reweigh the configurations thus sampled so as to recover the Boltzmann distribution.
One could quote here in this regard the Wang-Landau method\cite{Wang01}, the multicanonical ensemble\cite{Berg92}, the well-tempered ensemble\cite{Bonomi10}, nested sampling\cite{Partay10}, and integrated tempering sampling (ITS)\cite{Gao08}.
These approaches have two advantages, on the one hand they enhance the probability of escaping from the initial metastable state, on the other they allow computing the properties of the system at different temperatures in a single run. 
These methods are sometimes referred to as multicanonical ensembles and, of course, extension to multiple pressures is possible leading to multithermal-multibaric ensembles\cite{Okumura04,Shell02}. 

Here we shall use the variationally enhanced sampling (VES) \cite{Valsson14} method to obtain an efficient multithermal-multibaric sampling such that in a single simulation entire regions of the temperature-pressure plane can be explored.
We recall that in VES one introduces a functional of the bias $V(\mathbf{s})$:
\begin{align}
\label{omega1}
\Omega [V] & =
\frac{1}{\beta} \log
\frac
{\int d\mathbf{s} \, e^{-\beta \left[ F(\mathbf{s}) + V(\mathbf{s})\right]}}
{\int d\mathbf{s} \, e^{-\beta F(\mathbf{s})}}
+
\int d\mathbf{s} \, p(\mathbf{s}) V(\mathbf{s}),
\end{align}
where $\beta=1/k_BT$ is the inverse temperature, $\mathbf{s}$ is a set of collective variables (CVs) that are function of the atomic coordinates $\mathbf{R}$, the free energy is given within an immaterial constant by $F(\mathbf{s})=-\frac{1}{\beta}\log\int d\mathbf{R} \: \delta(\mathbf{s}-\mathbf{s}(\mathbf{R})) e^{-\beta U(\mathbf{R})}$,  $U(\mathbf{R})$ is the interatomic potential, and $p(\mathbf{s})$ is a preassigned target distribution.
The minimum of this convex functional is reached for
\begin{equation}
\label{eq:optimal_bias}
V(\mathbf{s}) = -F(\mathbf{s})-{\frac {1}{\beta}} \log {p(\mathbf{s})},
\end{equation}
which amounts to say that in a system biased by $V(\mathbf{s})$, the distribution is $p(\mathbf{s})$.   
The standard approach to solve the variational problem is to expand $V(\mathbf{s})$ is some basis set $f_i(\mathbf{s})$, such that
\begin{equation}
V(\mathbf{s})=\sum\limits_{i=1}^{N} \alpha_i f_i(\mathbf{s}) 
\end{equation}
where $\boldsymbol\alpha=(\alpha_1,...,\alpha_N)$ are variational coefficients that have to be determined, and $N$ is the order of the expansion.

Before discussing the multithermal-multibaric case, we shall deal with the simpler multicanonical scenario.
In the VES context it is relatively straightforward, at least conceptually, to design a multicanonical sampling.
One starts by choosing as CV the potential energy of the system as done, for instance, in the well-tempered ensemble.
We shall refer to this special CV as $E$, in order to underline its special role in statistical mechanics and distinguish it from more system-specific CVs.
Finally we impose a uniform sampling in the energy interval $E_1$--$E_2$ by choosing as the target distribution
\begin{equation}
p(E)= \begin{cases}
         \frac{1}{E_2-E_1} & \mathrm{if} \quad E_1<E<E_2 \\
         0 & \mathrm{otherwise}
      \end{cases}
      .
\label{eq:target_not_smooth}
\end{equation}

We now have to go back to our original task of determining the properties of the system at different temperatures.
Clearly the range $E_1$--$E_2$ chosen is related to the interval of temperatures $\beta_1$--$\beta_2$ in which the VES run conducted at temperature $\beta$ can be reweighted to give the properties of the system at a different temperature $\beta'$ with $\beta_1<\beta'<\beta_2$.
Although one could, in principle, first fix the interval $E_1$--$E_2$ and then determine $\beta_1$--$\beta_2$, in the spirit of this work we set up a predictor corrector procedure in which we use as input $\beta_1$--$\beta_2$ and later we determine the appropriate interval $E_1$--$E_2$.

The predictor corrector algorithm is based on a property of the free energy $F_{\beta}(E)$ in the canonical ensemble at temperature $\beta$.
Namely, $F_{\beta}(E)$ is simply related to the temperature independent density of states $N(E)$ by the relation\cite{Valsson13,TuckermanBook}:
\begin{equation}
F_{\beta}(E) =  E - \frac{1}{\beta} \log N(E) + C,
\end{equation}
where here we shall choose $C$ such that $F_{\beta}(E_{m})=0$ with $E_{m}$ the position of the free energy minimum.
If we consider two different temperatures $\beta$ and $\beta'$, and bearing in mind that the density of states is independent of temperature, we arrive at,
\begin{equation}
\label{eq:free_energy_other_temp}
\beta' F_{\beta'}(E) = \beta F_{\beta}(E) + (\beta' - \beta) E + C',
\end{equation}
where $C'$ is set by the relation $F_{\beta'}(E_{m}')=0$.
The last equation states that the free energy $F_{\beta'}(E)$ at temperature $\beta'$ can be easily calculated if $F_{\beta}(E)$ at temperature $\beta$ is known.
We will make use of Eq.\ \eqref{eq:free_energy_other_temp} to calculate $F_{\beta_1}(E)$ and $F_{\beta_2}(E)$ from our knowledge of $F_{\beta}(E)$.

\begin{figure}
\centering
\includegraphics[width=0.99\columnwidth]{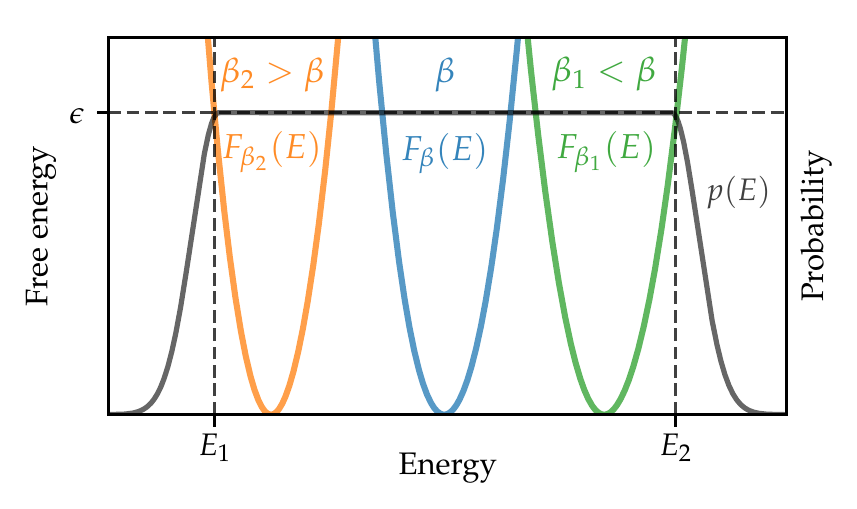}
\caption{\label{fig:Figure1} Illustration of the algorithm to find the energy interval $E_1$--$E_2$ from the temperature range $\beta_1$--$\beta_2$.
			     Free energies at temperatures $\beta$, $\beta_1$, and $\beta_2$ are shown in blue, green, and orange solid lines.
                             $p(E)$ is shown with a gray solid line.
                             The limits of the energy interval and the energy threshold $\epsilon$ are shown in black dashed lines.
                             }
\end{figure}

The scheme works as follows.
We first choose an initial guess for $E_1$ and $E_2$ that we shall call $E_1^0$ and $E_2^0$.
With these guessed values we begin the VES simulation using the target distribution $p^0(E)=\frac{1}{E_2^0-E_1^0}$ and get a first estimate $F_{\beta_1}^0(E)$ and $F_{\beta_2}^0(E)$ for $F_{\beta_1}(E)$ and $F_{\beta_2}(E)$.
$F_{\beta_1}^0(E)$ and $F_{\beta_2}^0(E)$ are calculated using Eq.\ \eqref{eq:free_energy_other_temp} with $F_{\beta}(E)$ obtained from Eq.\ \eqref{eq:optimal_bias}.
With this estimation we obtain a new value for $E_1$ by finding the leftmost solution of $\beta_2 F_{\beta_2}^0(E_1^1)=\epsilon$ where $\epsilon$ is a preassigned threshold and $E_1^1$ will be the estimation of $E_1$ at the first iteration.
Similarly the rightmost solution of  $\beta_1 F_{\beta_1}^0(E_2^1)=\epsilon$ gives us the estimation of $E_2$ at iteration 1.
In general, at iteration $k$ we have the estimates $E_1^k$ and $E_2^k$ for $E_1$ and $E_2$, respectively, and the target distribution becomes
\begin{equation}
p^k(E)=\frac{1}{E_2^k-E_1^k} \quad \mathrm{for} \quad E_1^k<E<E_2^k.
\end{equation}
$E_1^k$ and $E_2^k$ are obtained from $\beta_2 F_{\beta_2}^{k-1}(E_1^k)=\epsilon$ and $\beta_1 F_{\beta_1}^{k-1}(E_2^k)=\epsilon$.
The procedure is repeated until convergence.
This iterative approach is similar to that in Ref.\ \citenum{Valsson15}.
In Fig. \ref{fig:Figure1} we show a graphical interpretation of the scheme.
In the practice, instead of using the $p(E)$ described in Eq.\ \eqref{eq:target_not_smooth} we replace it by a smooth counterpart such as the one depicted in Fig. \ref{fig:Figure1}.
The example in this figure corresponds to a multicanonical, constant volume simulation of liquid Na between 400 K and 600 K.
This example, however, is rather simple and it is only discussed in the Supplemental Material.

In order to illustrate the fruitfulness of our approach, we set out to study the density anomaly in TIP4P/ice water\cite{Abascal05} in a single multithermal molecular dynamics simulation.
This water model has been extensively studied and the temperature of maximum density $T_{\mathrm{max}}$ at atmospheric pressure is 295 K\cite{Vega05} while the liquid-hexagonal ice coexistence temperature $T_m$ (or simply, melting temperature) at the same condition is 272.2 K.
Note that the difference between $T_m$ and $T_{\mathrm{max}}$ in this water model is $\sim23$ K while in real water is only $\sim4$ K.
Our simulation is performed at constant temperature 300 K and constant atmospheric pressure, and we wish to obtain information about temperatures from 260 to 350 K.
The same temperature range has been studied in Ref.\ \citenum{Vega05} using multiple isothermal-isobaric simulations.

Before discussing the results of our simulation we describe the computational details.
MD simulations of TIP4P/ice water\cite{Abascal05} were performed using Gromacs 2018.1\cite{Abraham15} patched with a development version of PLUMED 2\cite{Tribello14} supplemented by the VES module\cite{vescode}.
The electrostatic interaction in reciprocal space was calculated using the particle mesh Ewald (PME) method \cite{Essmann95}.
The atomic bonds involving hydrogen were constrained using the LINCS algorithm\cite{Hess08}.
The temperature was controlled using the stochastic velocity rescaling thermostat \cite{Bussi07} and the pressure was kept constant employing the isotropic version of the Parrinello-Rahman \cite{Parrinello81}.
MD simulations of Na (described in the Supplemental Material) were performed with LAMMPS\cite{Plimpton95} patched with PLUMED 2.
Na was described using an EAM potential\cite{Wilson15}.
Other details can be found in the Supplemental Material.

\begin{figure}
\centering
\includegraphics[width=0.99\columnwidth]{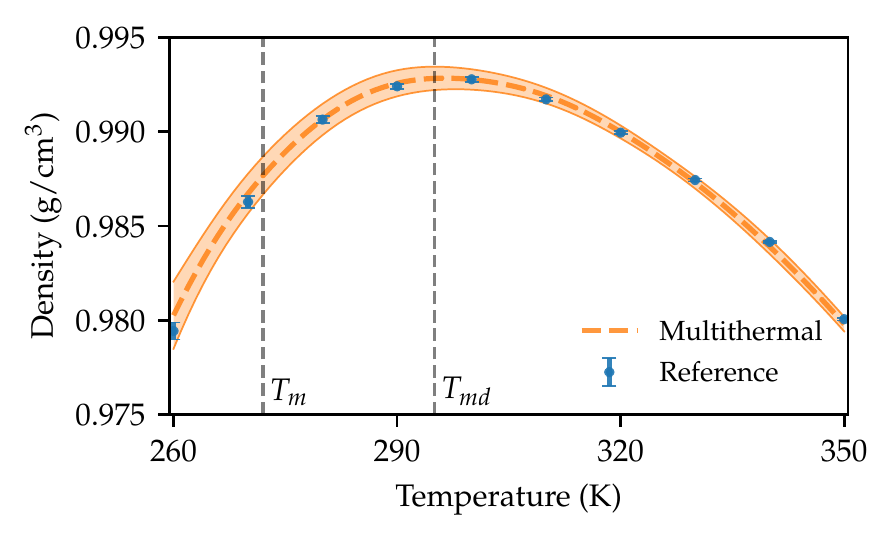}
\caption{\label{fig:Figure2} Density of TIP4P/ice liquid water at 1 bar as a function of temperature.
                             The blue circles with error bars are the results of individual isothermal-isobaric simulations.
                             Errors have been calculated using block averages\cite{Flyvbjerg89,FrenkelBook}.
                             The orange dashed line is the mean density calculated using Eq.\ \eqref{eq:reweight} from a single multithermal simulation.
                             The orange filled region is the error of the mean.
                             The melting temperature $T_m$ and the temperature of maximum density of the model $T_{\mathrm{max}}$ are also indicated.
                             }
\end{figure}

We now describe the results of our multithermal simulation that has a short transient of about 5 ns during which the coefficients $\boldsymbol\alpha$ are optimized\cite{Bach13} and the limits of the interval $E_1$--$E_2$ are determined (details are provided in the Supplemental Material).
After some degree of convergence is reached, the optimization is stopped and the simulation continued with fixed $\boldsymbol\alpha$ in order to gather statistics.
The simulation can then yield information at all temperatures in the interval from 260 to 350 K.
However, the simulation has been performed in a biased ensemble at temperature $\beta$ and in order to obtain properties of the isothermal-isobaric ensemble at temperature $\beta'$ each configuration must be properly weighed.
Basic statistical mechanics can be employed to calculate the mean value in the isothermal-isobaric ensemble at temperature $\beta'$ of an observable $O(\mathbf{R},\mathcal{V})$ that depends on the atomic coordinates $\mathbf{R}$ and the volume $\mathcal{V}$.
This is,
\begin{equation}
\langle O(\mathbf{R},\mathcal{V}) \rangle_{\beta'} = \frac{\langle O(\mathbf{R},\mathcal{V}) w(\mathbf{R},\mathcal{V}) \rangle_{\beta,V}}
                                                     {\langle w(\mathbf{R},\mathcal{V})  \rangle_{\beta,V}}
\label{eq:reweight}
\end{equation}
where $w(\mathbf{R},\mathcal{V})=e^{(\beta-\beta') [E(\mathbf{R})+P\mathcal{V}]} e^{\beta V(E)}$, $\langle \cdot \rangle_{\beta,V}$ is a mean value in the biased ensemble at temperature $\beta$ using a stationary bias potential $V(E)$, and $P$ is the pressure.
We employ Eq.\ \eqref{eq:reweight} to calculate the density as a function of temperature.
The results are shown in Fig. \ref{fig:Figure2} where they are also compared with individual isothermal-isobaric simulations.
The results are identical within the statistical error.
Fig. \ref{fig:Figure2} also highlights that the multithermal simulation provides continuous results as a function of temperature and there is no need to interpolate between different temperatures.
We note that Eq.\ \eqref{eq:free_energy_other_temp} is strictly valid only in constant volume simulations.
However, in this example the $PV$ term is very small and therefore changes in volume can be neglected.
In the next example we will present a more general but slightly more involved approach.
The version of the method in which a pressure interval is explored at constant temperature is discussed in the Supplemental Material together with an application to liquid Na.

\begin{figure}
\centering
\includegraphics[width=0.99\columnwidth]{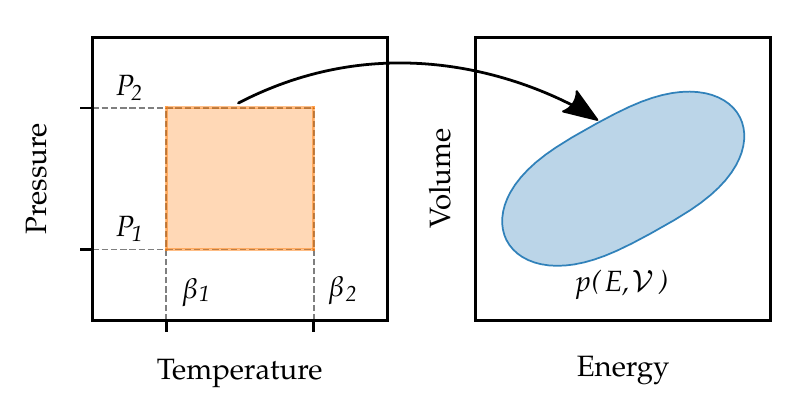}
\caption{\label{fig:Figure3} Idealized illustration of the mapping from a set of (inverse) temperatures and pressures to the set of relevant energies and volumes. 
                             The target distribution $p(E,\mathcal{V})$ corresponds to the latter set.
                             A realistic version of this plot is shown in the Supplemental Material.
                             }
\end{figure}

We now present the multithermal-multibaric generalization. 
Naturally, we shall use as collective variables in the variational principle the energy $E$ and the volume $\mathcal{V}$.
The starting point for the algorithm is an equation analogous to Eq.\ \eqref{eq:free_energy_other_temp}.
In the isothermal-isobaric ensemble the following equation holds,
\begin{align}
\beta' F_{\beta',P'}(E,\mathcal{V}) = & \beta F_{\beta,P}(E,\mathcal{V}) + (\beta' - \beta) E \nonumber \\ 
                            & + (\beta' P' - \beta P ) \mathcal{V} + C'',
\label{eq:free_energy_other_temp_press}
\end{align}
where $F_{\beta,P}(E,\mathcal{V})$ is the free energy as a function of the energy $E$ and volume $\mathcal{V}$ at temperature $\beta$ and pressure $P$, and $C''$ is a constant such that $\beta' F_{\beta',P'}(E_{m},\mathcal{V}_{m})=0$.
From this equation we shall construct the target distribution $p(E,\mathcal{V})$ needed in the variational principle.
The target distribution is defined as
\begin{equation}
p(E,\mathcal{V})=
  \begin{cases}
    1/\Omega_{E,\mathcal{V}} & \text{if there is at least one } \beta',P' \text{ such} \\
             & \text{that } \beta' F_{\beta',P'}(E,\mathcal{V})<\epsilon \text{ with}  \\
             & \beta_1<\beta'<\beta_2 \text{ and } P_1<P'<P_2 \\
    0 & \text{otherwise}
  \end{cases}
\end{equation}
where $\Omega_{E,\mathcal{V}}$ is a normalization constant, $\beta' F_{\beta',P'}(E,\mathcal{V})$ is calculated using Eq.\ \eqref{eq:free_energy_other_temp_press}, and $\epsilon$ is a predefined energy threshold.
$p(E,V)$ can be seen as a mapping from a set of temperatures and pressures to a set of energies and volumes.
This is illustrated in Fig. \ref{fig:Figure3}.
As in the multitemperature case, $F_{\beta,P}$ is not known beforehand and it is found during the minimization of $\Omega[V]$.
Therefore, also $p(E,\mathcal{V})$ is determined iteratively.
In simple terms the objective of the algorithm is to find a bias potential such that the final distribution of energy and volume contains the energies and volumes relevant at all the desired combinations of temperatures and pressures.

In order to provide an example of our method, we revisit TIP4P/ice water but this time we aim at studying the effect of temperature and pressure in the density anomaly in a single MD simulation.
We wish to study the same temperature interval as above (260--350 K) and the pressure interval 0--300 MPa.
The maximum pressure chosen here is approximately the pressure in the triple point liquid-ice Ih-ice III (see the phase diagram in Ref.\ \citenum{Abascal05}).
The simulation takes around 50 ns to converge.
The convergence of the coefficients $\boldsymbol\alpha$ and $p(s)$ is discussed in the Supplemental Material.
It is also important to assess whether there is overlap between the unbiased distributions of $E$ and $\mathcal{V}$, and the biased ones.
This analysis is presented in the Supplemental Material and shows that the relevant region of energy and volume is identified with great accuracy.
Therefore, the method guarantees an economical sampling of the chosen $TP$ region in which no time is wasted in visiting irrelevant regions.
Once that the $\boldsymbol\alpha$ are converged, we continue the simulation for 200 ns with fixed $\boldsymbol\alpha$.
In the next paragraphs we illustrate the surprising amount of information that can be extracted from this simulation.

\begin{figure}
\centering
\includegraphics[width=0.99\columnwidth]{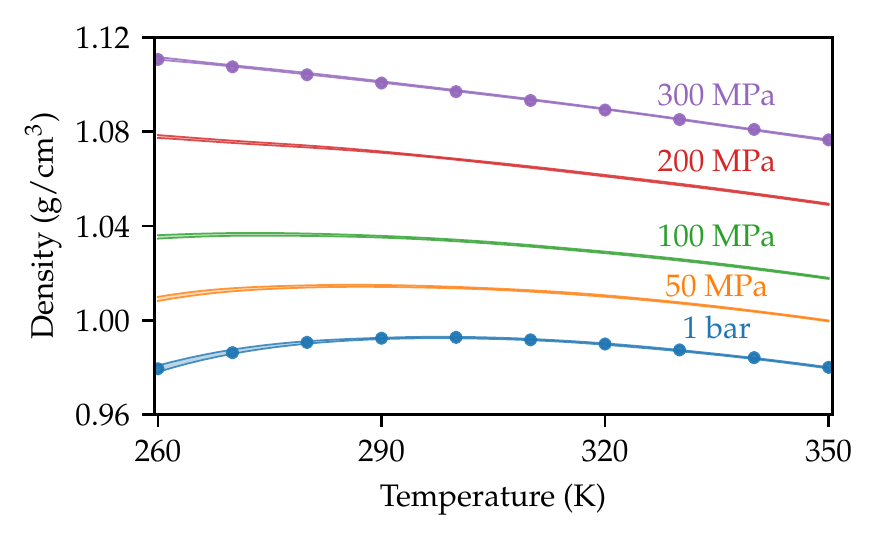}
\caption{\label{fig:Figure4} Density of TIP4P/ice liquid water as a function of temperature for different pressures.
                             The solid lines are the result from a single multithermal-multibaric simulation.
			     The circles are references from individual isothermal-isobaric simulations at 1 bar and 300 MPa.
                             }
\end{figure}

\begin{figure}[b]
\centering
\includegraphics[width=0.99\columnwidth]{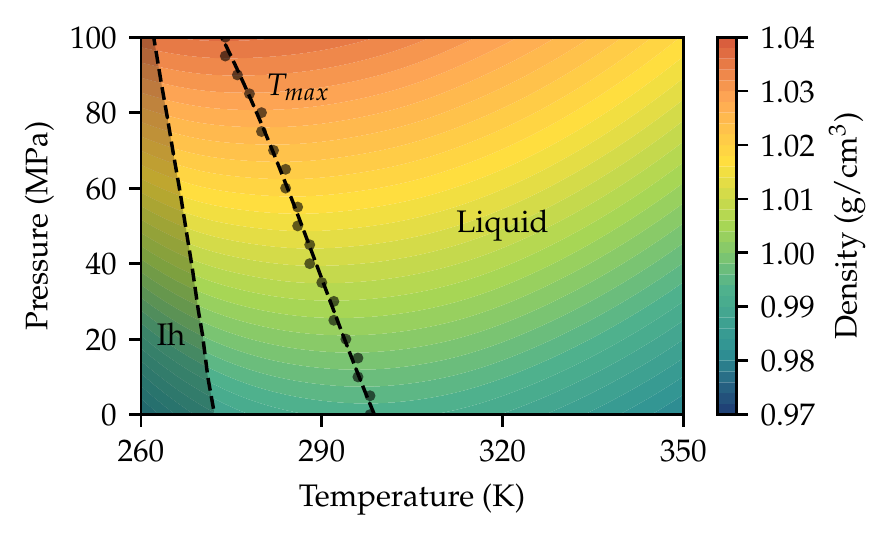}
\caption{\label{fig:Figure5} Density of TIP4P/ice liquid water as a function of temperature and pressures.
                             The temperature of maximum density $T_{\mathrm{max}}$ for different pressures is shown in black circles.
                             $T_{\mathrm{max}}$ is also represented using a $B$-spline approximation to the points.
                             The region were liquid water is undercooled with respect to ice Ih has been shaded in gray.
			     We stress that only the liquid state was sampled in the simulation.
                             }
\end{figure}

As in the multitemperature case, the mean value of an observable in the isothermal-isobaric ensemble at temperature $\beta'$ and pressure $P'$ can be calculated from the multithermal-multibaric simulation using,
\begin{equation}
\langle O(\mathbf{R},\mathcal{V}) \rangle_{\beta',P'} = \frac{ \langle O(\mathbf{R},\mathcal{V}) w(\mathbf{R},\mathcal{V}) \rangle_{\beta,P,V}}
                                                             { \langle w(\mathbf{R},\mathcal{V}) \rangle_{\beta,P,V}},
\label{eq:reweight2}
\end{equation}
where $w(\mathbf{R},\mathcal{V})=e^{(\beta-\beta')E(\mathbf{R}) + (\beta P - \beta' P') \mathcal{V}} e^{\beta V(E)}$.
We will apply this formula to calculate different observables.
We start by calculating the density as a function of temperature and pressure.
In Fig. \ref{fig:Figure4} we plot the density as a function of temperature at different pressures.
We compare our results with individual isothermal-isobaric simulations for the pressures 1 bar and 300 MPa (circles in Fig. \ref{fig:Figure4}) and the agreement is excellent.
Analyzing isobars in Fig. \ref{fig:Figure4} allowed us to compare our results.
However, our simulation provides continuous information in temperature and pressure.
Thus in Fig. \ref{fig:Figure5} we show a contour plot of the density as a function of temperature and pressure.
From this data we can also calculate $T_{\mathrm{max}}$ for different pressures (black circles in Fig. \ref{fig:Figure5}).
This example highlights that relevant thermodynamic information as a function of temperature and pressure can be obtain from only one simulation.

\begin{figure}
\centering
\includegraphics[width=0.99\columnwidth]{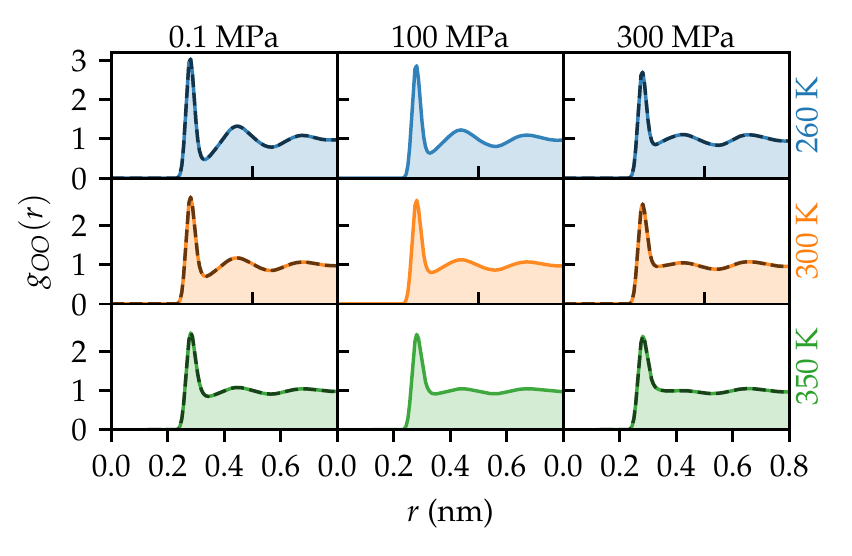}
\caption{\label{fig:Figure6} Oxygen-oxygen radial distribution functions $g_{OO}(r)$ of TIP4P/ice liquid water for different temperatures and pressures.
                             Results from the multithermal-multibaric simulations are shown in solid blue, orange, and green lines for the temperatures of 260, 300 and 350 K, respectively.
                             The columns correspond to pressures of 0.1 MPa (1bar), 100 MPa, and 300 MPa. 
                             References calculated during individual isothermal-isobaric simulations are shown in dashed gray lines.
                             }
\end{figure}

The anomalous properties of water, for instance, the density maximum in the liquid, are a result of its structure.
It is therefore interesting to characterize the structure of water as a function of temperature and pressure.
One way to do so is by quantifying the tetrahedral order around each water molecule.
In the Supplemental Material we calculate the tetrahedral order parameter defined in Ref.\ \citenum{Errington01} as a function of temperature and pressure.
Another way to study the structure of liquids is by calculating the radial distribution function $g(r)$.
From our multithermal-multibaric simulation we calculated using Eq.\ \eqref{eq:reweight2} the oxygen-oxygen radial distribution function $g_{OO}(r)$ for different temperatures and pressures (see Fig. \ref{fig:Figure6}).
The agreement of our results with the $g_{OO}(r)$ calculated using individual isothermal-isobaric simulations is very good.
In this way, from a single simulation we can observe that water becomes less structured as the temperature and pressure increase.

In this example we have chosen a region of the phase diagram in which there are no first-order phase transitions.
If this was the case, similarly to what has been done in Ref.\ \citenum{Yang18}, one should combine our approach with some collective variable based enhanced sampling method such as metadynamics\cite{Laio02}.
This extension will be discussed elsewhere.

The computational advantages of the method are obvious since the global cost for the whole $TP$ plane in the case of water is $\sim$ 300 ns.
This is to be compared with the cost of a single $TP$ point calculation in which the error bar are comparable to ours, i.e. $\sim$ 5 ns.
We stress that in our approach dynamical properties cannot be computed.
The method developed here has been implemented in the PLUMED 2 enhanced sampling plugin\cite{Tribello14} that can be interfaced with most of the popular \textit{ab initio} and classical MD codes.
In the present work we performed our calculations with LAMMPS and Gromacs showing the versatility of our implementation.
We plan to make these tools available to the community in the near future.

\begin{acknowledgments}
P.M.P would like to thank Mario Del Popolo for insightful discussions concerning the connection between different methods to perform multitemperature simulations.
M.P thanks Vanda Glezakou for useful discussions.
We would also like to thank Yi Isaac Yang for sparking our interest in multitemperature simulations.
This research was supported by the NCCR MARVEL, funded by the Swiss National Science Foundation.
The authors also acknowledge funding from European Union Grant No. ERC-2014-AdG-670227/VARMET.
The computational time for this work was provided by the Swiss National Supercomputing Center (CSCS) under Project ID mr22.
Calculations were performed in CSCS cluster Piz Daint.
\end{acknowledgments}

\end{document}